\documentclass[preprint,showpacs,amsmath,amssymb,graphicx]{revtex4}
\usepackage[dvips]{graphicx}
\begin{document}
\def\bea{\begin{eqnarray}}
\def\eea{\end{eqnarray}}
\def\nn{\nonumber}
\newcommand{\snu}{\tilde \nu}
\newcommand{\sll}{\tilde{l}}
\newcommand{\asnu}{\bar{\tilde \nu}}
\newcommand{\stau}{\tilde \tau}
\newcommand{\dmsnu}{{\mbox{$\Delta m_{\tilde \nu}$}}}
\newcommand{\mt}{{\mbox{$\tilde m$}}}

\renewcommand\epsilon{\varepsilon}
\def\beq{\begin{equation}}
\def\eeq{\end{equation}}
\def\lla{\left\langle}
\def\rra{\right\rangle}
\def\za{\alpha}
\def\zb{\beta}
\def\lsim{\mathrel{\raise.3ex\hbox{$<$\kern-.75em\lower1ex\hbox{$\sim$}}} }
\def\gsim{\mathrel{\raise.3ex\hbox{$>$\kern-.75em\lower1ex\hbox{$\sim$}}} }
\newcommand{\Rbs}{\mbox{${{\scriptstyle \not}{\scriptscriptstyle R}}$}}

\draft
\preprint{{\vbox\hbox{CERN-PH-TH/2005-077}{\hbox{SNUTP-05-005}
\hbox{May 2005} }}}

\title{Sneutrino Leptogenesis at the Electroweak Scale}

\thispagestyle{empty}
\author{ John Ellis
\thanks{ E-mail : John.Ellis@cern.ch}}
\address{\small \it TH Division, PH Department, CERN, 1211 Geneva
23, Switzerland }
\author{ Sin Kyu Kang
\thanks{ E-mail : skkang@phya.snu.ac.kr}}
\address{\small \it School of Physics, Seoul National University,
       Seoul 151-734, Korea }

\begin{abstract}

We propose an alternative mechanism for leptogenesis at the electroweak
scale, through the decays of a left-handed sneutrino. This scenario may be
realized in supersymmetric models with non-zero Majorana masses for the
neutrino superfield that lead to mixing and mass splitting between the
left-handed sneutrino and the corresponding antisneutrino. Soft
supersymmetry breaking provides new sources of CP violation in
sneutrino-antisneutrino mixing that can generate a lepton asymmetry in
decays of the left-handed sneutrino. We show how the three Sakharov
conditions for generating the observed baryon asymmetry of the Universe
can be fulfilled in this restrictive framework.
~~\\
\begin{flushleft}
CERN-PH-TH/2005-077, SNUTP-05-005\\
May 2005
\end{flushleft}

\end{abstract}
\pacs{11.30.Fs, 12.60.Jv,  14.80.Ly, 98.80Cq \\
Keywords: leptogenesis, supersymmetry breaking, sneutrino mixing}
 \maketitle \thispagestyle{empty}
%

\section{Introduction}

The baryon asymmetry of the universe may be expressed in terms of
the ratio of the baryon density $n_B$ to the entropy density $s$
of the Universe. The agreement between astrophysical observations
and primordial nucleosynthesis calculations constrains this ratio
to the range at $95\%$ C.L.~\cite{pdg}:
\begin{equation}
\frac{n_B}{s} \; \simeq \; (3.4 - 6.9) \cdot 10^{-10}, \label{nbs}
\end{equation}
which is in good agreement with the inferred from Cosmic Microwave
Background data, particularly that provided by WMAP~\cite{cmb}.
Almost 40 years ago, Sakharov explained how this baryon asymmetry could be obtained
via a microphysical mechanism incorporating interactions that violate
baryon and/or lepton number, violate C and CP, and drop out of equilibrium
in the early Universe~\cite{sakharov}.

Since such interactions are present in the electroweak theory and its
extensions, the baryon asymmetry might have originated at any energy above
the electroweak scale. One of the most attractive and simple ways to
realize the Sakharov mechanism is high-scale leptogenesis  through the
out-of-equilibrium decays of a heavy Majorana neutrino,
since it arises naturally within the seesaw mechanism for
generating light neutrino masses \cite{lepto}. However, high-scale
leptogenesis and the seesaw mechanism are difficult to test
directly, since the natural scale of the interactions of the heavy
Majorana neutrino is in the range $10^{10}$ GeV to $10^{15}$ GeV.
This is impossible to reach by accelerators, except indirectly via
renormalization effects on light particle masses, for example.

Baryogenesis at the electroweak scale itself is certainly a most
attractive option\cite{electbary}, because it may be tested
directly, if it can be achieved economically.
However, this option is now excluded in the Standard Model,
because the unsuccessful LEP searches require the Higgs boson to be
too heavy for the electroweak phase transition to have been first-order,
and the effective amount of CP violation is in any case very small
\cite{smphase}. On the other hand, an electroweak mechanism may be
possible in the minimal supersymmetric extension of the Standard
Model (MSSM), but at the price of a certain number of rather
restrictive conditions \cite{susyphase}. It is therefore
worthwhile to look for other scenarios for electroweak
baryogenesis.

An additional problem with any supersymmetric model that
postulates high temperatures in the early Universe is that of the
gravitino abundance \cite{gravitino}.
Thermal production of gravitinos at high temperatures in the very early
Universe might have exceeded the bounds imposed by primordial
nucleosynthesis and the Cosmic Microwave Background \cite{reheat}.
Therefore, in the context of supersymmetry, it is doubly interesting
to look for alternative baryogenesis scenarios that operate at lower energy
scales below about a TeV, which might be tested directly in the
foreseeable future.

It has recently been proposed that soft supersymmetry-breaking breaking
terms comprising bilinear and trilinear scalar couplings involving the
right-handed sneutrino fields could be responsible for leptogenesis, an
option referred to as soft leptogenesis~\cite{softl}.
In contrast to standard leptogenesis from heavy Majorana (s)neutrino decay,
in soft leptogenesis the CP violation needed can be provided within the framework
of a single sneutrino generation. However, even if this mechanism of
leptogenesis could evade the gravitino problem, it is still difficult to
probe in laboratory experiments, because of the high mass scale of the
right-handed sneutrino.

In this paper, we propose an alternative leptogenesis mechanism via {\it
decays of a left-handed sneutrino at a low energy scale}.
This scenario may be realized in supersymmetric models with non-zero Majorana
neutrino masses that lead to mixing and mass splitting between the left-handed
sneutrino $\snu$ and the corresponding antisneutrino $\asnu$.
As was shown in~\cite{grossman, smixing}, the neutrino mass and the sneutrino mass
splitting may be due to related $\Delta L=2$ interactions, linked by
supersymmetry breaking. As we show below, the $\snu-\asnu$ system may
exhibit oscillatory behaviour analogous to the $B-\bar{B}$ system.
In addition, soft supersymmetry breaking sectors provide new sources of CP
violation in $\snu$-$\asnu$ mixing which can generate a lepton asymmetry
in the sneutrino decays. We explore in this paper how the three Sakharov
conditions~\cite{sakharov} for generating the observed baryon asymmetry of
the Universe can be fulfilled in this restrictive framework.

In the sense that supersymmetry breaking is the source of
leptogenesis, our scenario is similar to soft leptogenesis, but
differs in that it operates at the electroweak scale. If the
scenario proposed does become a candidate for leptogenesis, it
could in principle be tested in collider experiments via a
sneutrino oscillation signal, though this may be difficult in
practice. The lepton number is tagged in the decay of sneutrino by
identifying the charge of the outgoing lepton, and a same-sign
dilepton signal may be observable when the sneutrino-antisneutrino
pairs decay into charged leptons. However, measurement of a lepton
asymmetry would probably need more events than are likely to be
provided at the LHC or ILC.

The outline of this paper is as follows. In the next section, we review
how to generate the sneutrino mass splitting in a supersymmetric model
with a heavy right-handed Majorana neutrino superfield and discuss how
small the mass splitting can be. In section III, we examine how
leptogenesis via the decay of the left-handed sneutrino can be realized
and in section IV the CP asymmetry generated from the decay of the light
sneutrino is considered. In section V, we discuss possible wash-out
processes, which may be harmless for our leptogenesis scenario, and
present our conclusions.

\section{Light Sneutrino Masses}

We first review in more detail how to generate the
sneutrino mass splitting in a supersymmetric model with non-zero
Majorana neutrino masses, which could arise from a right-handed neutrino
superfield. The relevant superpotential is
\begin{eqnarray}
W=Y_{\nu} \widehat{H}_2 \widehat{L}\widehat{N}-\mu
\widehat{H}_1\widehat{H}_2+\frac{1}{2} M \widehat{N}\widehat{N} ,
\label{superpotential}
\end{eqnarray}
where $\widehat{L},\widehat{N},\widehat{H}_i(i=1,2)$ are
superfields for lepton doublets, heavy right-handed neutrinos and
Higgs fields, respectively. The $D$ terms are the same as in the
MSSM. The relevant terms in the soft supersymmetry-breaking scalar
potential are given by
\begin{eqnarray}
V_{soft}=m^2_{\widetilde{L}}\widetilde{\nu}^{\ast}\widetilde{\nu}
+m^2_{\widetilde{N}}\widetilde{N}^{\ast}\widetilde{N} +(Y_{\nu}
A_NH_2\widetilde{\nu}\widetilde{N}^{\ast}+MB_N\widetilde{N}\widetilde{N}
+H.c.) ,
\end{eqnarray}
where $\widetilde{\nu}$ and $\widetilde{N}$ are the light and
heavy sneutrinos, respectively.  From now on, we consider a single
generation of $\widehat{L}$ and $\widehat{N}$. After the
electroweak symmetry is spontaneously broken, the light neutrino
acquires a mass via the seesaw mechanism: $m_{\nu}\simeq m^2_D/M$,
where $m_D=Y_{\nu}v_2$ with $v_2/\sqrt{2}=\langle H_2^0\rangle$.
Defining
\begin{eqnarray}
\widetilde{\nu}_1 &=&
(e^{i\phi/2}\widetilde{\nu}+e^{-i\phi/2}\widetilde{\nu^{\ast}})/\sqrt{2},~~~~~~
\widetilde{\nu}_2=-i(e^{i\phi/2}\widetilde{\nu}-e^{-i\phi/2}\widetilde{\nu}^{\ast})/\sqrt{2},
\nonumber \\
\widetilde{N}_1 &=&
(e^{i\phi^{\prime}/2}\widetilde{N}+e^{-i\phi^{\prime}/2}\widetilde{N^{\ast}})/\sqrt{2},~~~~
\widetilde{N}_2=-i(e^{i\phi^{\prime}/2}\widetilde{N}-e^{-i\phi^{\prime}/2}\widetilde{N^{\ast}})/\sqrt{2},
\end{eqnarray}
one can separate the sneutrino mass-squared matrix
into two  blocks and then, to first order in $1/M$,
the two light sneutrino eigenstates are $\widetilde{\nu}_1$ and
$\widetilde{\nu}_2$, with the following masses-squared,
\begin{eqnarray}
m^2_{\widetilde{\nu}_{1,2}}=m^2_{\widetilde{L}}+\frac{1}{2}m^2_Z\cos2\beta\mp
\frac{1}{2}\Delta m^2_{\widetilde{\nu}} ,
\end{eqnarray}
where the mass-squared difference $\Delta
m^2_{\widetilde{\nu}}=m^2_{\widetilde{\nu}_2} -
m^2_{\widetilde{\nu}_1} $ is of order $1/M$.
Here, we can remove the phases in the superpotential
(\ref{superpotential}) by rotating superfields and the relative
phase between $A_N$ and $B_N$ with a R-rotation, but there remains
an unremovable phase, which makes $\Delta m^2_{\snu}$ complex.
The sneutrino mass splitting is then easily computed by using the
relation $\Delta m^2_{\widetilde{\nu}}=2m_{\widetilde{\nu}}\Delta
m_{\widetilde{\nu}}$ \cite{grossman},
where
$m_{\widetilde{\nu}}=\frac{1}{2}(m_{\widetilde{\nu}_1}+m_{\widetilde{\nu}_2})$
is the average of the light sneutrino masses and
\begin{eqnarray}
\Delta m_{\widetilde{\nu}}\simeq \frac{2 m_{\nu}(A_N-\mu\cot
\beta-B_N)}{m_{\widetilde{\nu}}}.
\label{sp1}
\end{eqnarray}
We note that $\mu, A_N$ and $m_{\widetilde{L}}$ are of the same
order as the electroweak scale, whereas $M, m_{\widetilde{N}}$ and
$B_N$ are soft-supersymmetry breaking parameters associated with
the $SU(2)\times U(1)$ singlet superfield $\widehat{N}$, and may
be much larger than the electroweak scale. As one can see from
(\ref{sp1}), if $B_N >> m_Z$, the sneutrino mass splitting is significantly enhanced,
whereas it is of the same order as the neutrino mass when $B_N\sim
{\cal O}(m_Z)$. This mass splitting may in principle be probed through the
sneutrino-antisneutrino oscillation which would result in a
same-sign dilepton signal. However, to have an observable rate for the
same-sign dilepton signal, the ratio of the mass splitting to the
sneutrino decay width should be large, namely, $\Delta
m_{\snu}/\Gamma_{\snu}\geq 1$, and the sneutrino branching ratio
into a charged lepton should also be large.

\section{Conditions for Leptogenesis}

We now examine how leptogenesis via the decay of the light sneutrino can
be realized in this framework. As is well known, for a successful
mechanism of leptogenesis, we need lepton-number- and CP-violating
interactions that should be out of equilibrium, so that the asymmetry
generated is not automatically suppressed. In this scenario, as mentioned
before, {\it sneutrino-antisneutrino mixing is a direct manifestation of
lepton-number violation}, and the {\it soft supersymmetry-breaking terms
may well provide a suitable new source of CP violation}.

We must then impose the {\it out-of-equilibrium condition for
the decay of the light sneutrino}, which is given by
\begin{eqnarray}
\Gamma_{\widetilde{\nu}} < H \simeq
1.7\sqrt{g_{\ast}}\frac{T^2}{M_P}, ~~~~~\mbox{at} ~
T=m_{\widetilde{\nu}},
\label{out1}
\end{eqnarray}
where $\Gamma_{\widetilde{\nu}}$ is the decay rate of the
sneutrino, $H$ is the Hubble constant at the decay epoch, and $M_P$ is the
Planck scale. The parameter $g_{\ast}$ is the effective number of
massless degrees of freedom, which takes the value $g_{\ast}=225$ for
the MSSM with one generation of right-handed neutrinos.

As well as the out-of-equilibrium condition (\ref{out1}), we must
require that the electroweak {\it sphaleron interactions are still in
thermal equilibrium} at the time the lepton asymmetry is generated, so
that they can convert the lepton asymmetry partially into a baryon
asymmetry of the Universe. The temperature at which the sphaleron
interactions freeze out depends on how the electroweak phase
transition occurs. It is known that the sphaleron interactions
freeze out at the critical temperature of electroweak phase
transition if it is strongly first-order, whereas the freeze-out
temperature may become lower than the critical temperature if the
transition is second-order or weakly first-order.
In the latter case, the sphaleron interactions freeze out at the temperature at
which the sphaleron transition rate $\Gamma_{sph}$ becomes equal
to the expansion rate of the Universe\cite{sph}. It has been found
in numerical simulations that the sphaleron interactions are
effective as long as $T\geq 200$ GeV. Therefore, for a successful
mechanism of leptogenesis, we require
\bea
5\times
10^{-4}~\mbox{eV} \lsim \Gamma_{\snu} \lsim 1.3\times
10^{-4}\left(\frac{m_{\snu}(\mbox{GeV})}{100
\mbox{GeV}}\right)^2~\mbox{eV} ,
\label{out2}
\eea
where the first
condition follows from the sphaleron equilibrium condition and the
second from (\ref{out1}) for $T=m_{\snu}$.
As seen in (\ref{out2}), a scenario for leptogenesis can be successful only
when $m_{\snu}\gsim 200$ GeV.

\section{CP-Violating Sneutrino Decays and Mixing}

In order to discuss the decays, mixing and CP-violating asymmetries for
light sneutrinos, we first integrate out the heavy right-handed neutrino
superfield, following which the superpotential for the light sneutrino is:
\begin{eqnarray}
W=\lambda_{ij}\widehat{\nu_i}\widehat{l^c}_{Rj}\widehat{H_1}+\frac{\kappa_{ij}}{M}\widehat{H_2}
\widehat{\nu_i}\widehat{\nu_j}\widehat{H^{T}_2} ,
\label{seesaw2}
\end{eqnarray}
where $\widehat{\snu},\widehat{l_R^c},\widehat{H_1}$ denote the
light neutrino, the right-handed charged lepton and the charged
Higgs superfields, respectively, and $\lambda_{ij}$ stands for the
Yukawa couplings of the lepton sector, which are given by
$\lambda_{ij}=-gm_{l_{ij}}/\sqrt{2}M_W\cos\beta$.
The soft supersymmetry-breaking terms involving the light sneutrinos
$\widetilde{\nu}$ are;
\begin{eqnarray}
-L_{soft}=m^2_{\widetilde{L}}\widetilde{\nu}^{\ast}\widetilde{\nu}+\Delta
m^2_{\widetilde{\nu}}\widetilde{\nu}\widetilde{\nu}+\lambda
A_{\nu}\widetilde{\nu}\widetilde{l_R}H_1 .
\end{eqnarray}
We note that the effects of the right-handed neutrino superfields
are absorbed into the seesaw term in (\ref{seesaw2}) and the
sneutrino mass splitting $\Delta m^2_{\snu}$ discussed previously.
The sneutrino interaction Lagrangian is then given by
\begin{eqnarray}
L &=& \widetilde{\nu}(\lambda\overline{\widetilde{H_1}}l_R+
m_{\nu}\lambda^{\ast}\widetilde{l}^{\ast}_{Ri}H^{\ast}_1
+A_{\nu}\lambda\widetilde{l}_R H_1
+g^{\prime}Z_{iZ}\widetilde{\chi}_i^0\nu
+gV_{11}\widetilde{\chi}^+l^-_L)+h.c. ,
\label{L}
\end{eqnarray}
where we have considered a single sneutrino generation and
$m_{\nu}=\kappa v^2_2/M$ after the electroweak symmetry is broken.

It should be noted that the Higgs field $H_1$ is decomposed into the
physical Higgs sector and the Goldstone boson sector. Selecting the
physical Higgs sector in $H_1$ is equivalent to replacing $H_1$ by
$H^{-}\sin\beta$. Its Yukawa coupling therefore becomes
$-gm_l\tan\beta/\sqrt{2}M_W$, where $m_l$ is the charged lepton mass. In
the Lagrangian for a single generation of sneutrino, there is a physical
CP-violating phase. With superfield rotations and an R-rotation we can
eliminate CP phases in the parameters $\Delta m^2_{\snu}$, $\lambda$ and
$m_{\nu}$, but then the CP-violating phase in $A_{\nu}$ cannot be removed.

As observed above in (\ref{out2}), the mechanism of low-energy
leptogenesis proposed here could be successful only if $\Gamma_{\snu}$ is
of order $10^{-4}$~eV, so it is important to discuss how such a low decay
rate might be achieved.
If $m_{\tilde{\chi_1}^0} < m_{\tilde{\chi}^+},
m_{\snu}$, which is the ordering of masses generally expected in the MSSM
with universal soft supersymmetry-breaking scalar masses (the CMSSM), the
dominant sneutrino decays are those into two-body final states. As shown
in~\cite{grossman}, the typical size of $\Gamma_{\snu}$ for such two-body
decays is ${\cal O}$(GeV), which is too large to achieve successful
leptogenesis.

\begin{figure}
  \includegraphics{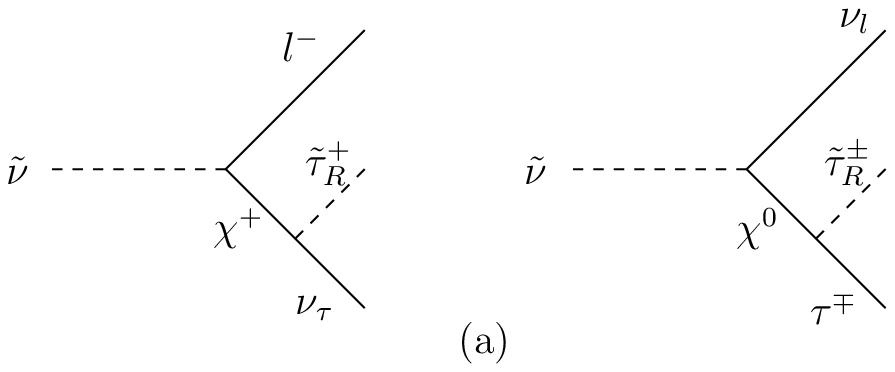}
  \includegraphics{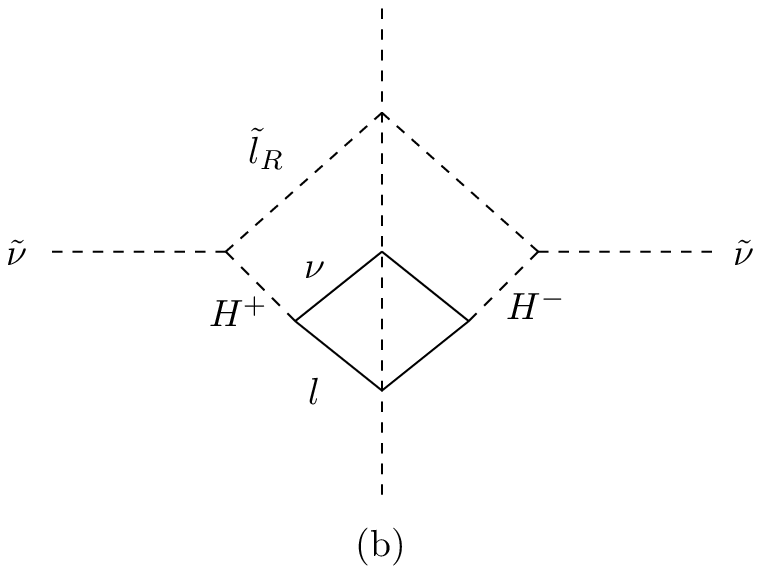}
    \caption{
    {\it (a) Diagrams for three-body sneutrino decay via charginos and
neutralinos, and (b) a diagram contributing to $\Gamma_{12}$. }
    }
  \label{fig:threebody}
\end{figure}

On the other hand, if $m_{\snu} < m_{\tilde{\chi_1}^0},m_{\tilde{\chi}^+}$
and no two-body sneutrino decay channels are open,
three-body sneutrino decays will dominate.
This mass ordering is possible if the gravitino is the
lightest supersymmetric particle, and is also allowed by cosmology
and the standard accelerator constraints~\cite{ellis}.
The following chargino- and neutralino-mediated three-body decays are
then generally dominant: $\snu_l\rightarrow
l^-\tilde{\tau}^+_R\nu_{\tau}$ and $\snu_l\rightarrow
\nu_l\tilde{\tau}^{\pm}\tau^{\mp}$, assuming that decays into
final states containing lighter sneutrinos can be neglected.
Assuming that the lightest neutralino is dominated by its bino
component, the rates for the chargino- and neutralino-mediated
sneutrino decays (see Fig.~\ref{fig:threebody}(a) for an
illustration) are given by \cite{grossman}
\bea
\Gamma(\snu_l\rightarrow l^-\tilde{\tau}^+_R\nu_{\tau}) &=&
       \frac{g^4m^3_{\snu}m^2_{\tau}\tan^2\beta
       f_c(m^2_{\stau}/m^2_{\snu})}{3\times 2^9\pi^3
       (m_W^2\sin2\beta -M_2\mu)^2}, \\
\Gamma(\snu_l\rightarrow \nu_l\tilde{\tau}_R^{\pm}\tau^{\mp}) &=&
\frac{g^{\prime 4}m^5_{\snu}f_n(m^2_{\stau_R}/m^2_{\snu})}{3\times
2^{10}\pi^3M_1^4},
\eea
where the $M_i$ are gaugino mass parameters,
$f_c(x) \equiv (1-x)(1+10x+x^2)+3x(1+x)\log x^2$ and $f_n(x) \equiv
1-8x+8x^3-x^4-6x^2\log x^2$.
%
%
%

In order to obtain $\Gamma\simeq 5\times 10^{-4}$~eV, one must
require that the $\snu$ and $\stau_R$ be nearly degenerate.
As an example, for $m^2_{\stau_R}/m^2_{\snu}=0.95$,
$m_{\snu}\simeq 500 ~\mbox{GeV},$ and $M_1\simeq 350 ~\mbox{GeV}$,
we indeed find $\Gamma\simeq 5\times 10^{-4}$ eV.
This near-degeneracy between the $\stau_R$ and $\snu$ is not possible
in the CMSSM with universal soft scalar masses $m_0$ and
gaugino masses $m_{1/2}$ at the GUT scale,
in which the mass difference between the $\stau_R$
and $\snu$ is approximately $m^2_{\snu}-m^2_{\stau_R}\simeq
0.4m^2_{1/2}+0.27\cos2\beta m_Z^2$.
According to the constraints on the parameter space $(m_0,m_{1/2})$
from the cosmological observation,
$b\rightarrow s \gamma$ and muon $g-2$, the favoured
region of $m_{1/2}$ is somewhat large, namely $m_{1/2}\geq 300 $
GeV, and the $\stau_R$ is considerably lighter than the $\snu$
\cite{ellis} in the CMSSM. Thus, we need a non-universal boundary
condition on the soft scalar masses at the GUT scale.
There is no known theoretical or phenomenological
reason why the soft superymmetry-breaking $\stau_R$ and $\snu$
masses should not differ at the GUT scale. At the electroweak
scale, $m_{\stau_R}$ and $m_{\snu}$ can be written as \bea
m_{\stau_R} &\simeq&
m_{0R}^2+0.15m_{1/2}^2+\sin^2\theta_W\cos2\beta m_Z^2, \\
m_{\snu} &=& m_{0L}^2+0.54m_{1/2}^2+1/2\cos2\beta m_Z^2 ,
\eea
and we can make the physical $m_{\stau_R}$ and $m_{\snu}$ nearly
degenerate by choosing the soft scalar masses $m_{0R}$ and $m_{0L}$
appropriately.

We now consider mass mixing and CP violation in the
$\widetilde{\nu}-\widetilde{\nu}^{\ast}$ system.
Since it is analogous to the $K^0-\overline{K}^0$ and $B^0-\overline{B}^0$
systems, we can calculate mixing and CP asymmetries using the
formulae derived previously for these meson systems \cite{bphys}.
The evolution of the system is determined by a Hamiltonian
$H=\widehat{M}-i\widehat{\Gamma}/2$ where, to leading order in the
soft terms,
\begin{eqnarray}
\widehat{M} &=& m_{\widetilde{\nu}}\left(%
\begin{array}{cc}
  1 &  \frac{\Delta m_{\widetilde{\nu}}}{m_{\widetilde{\nu}}}\\
  \frac{\Delta m_{\widetilde{\nu}}}{m_{\widetilde{\nu}}} &  1\\
\end{array}%
\right) ,\\
\widehat{\Gamma} &=& \left( \begin{array}{cc}
   \Gamma & \Gamma_{12} \\
   \Gamma_{12}^{\ast} & \Gamma , \\
\end{array}%
\right)
\end{eqnarray}
where $\Gamma$ is the total $\widetilde{\nu}$ decay width.
The dominant part of the off-diagonal
component $\Gamma_{12}$ can be obtained by considering the imaginary
part of the loop diagram shown in Fig.~\ref{fig:threebody}(b).
The analytic expression for $\Gamma_{12}$ is given by
\begin{eqnarray}
\Gamma_{12}=\frac{g^4\tan^4\beta m_l^4
m_{\nu}A_{\nu}}{512\pi^3M_W^4m^3_{\snu}}I
\end{eqnarray}
where
\begin{eqnarray}
I=\int^{(m_{\snu}-m_{\sll})^2}_0 ds
\frac{s\sqrt{s^2-2(m^2_{\snu}+m_{\sll}^2)s+(m_{\snu}^2-m_{\sll}^2)^2}}
{(s-m^2_{H^{-}})}.
\end{eqnarray}
The eigenvectors of the Hamiltonian $H$ are then given by
\begin{eqnarray}
\widetilde{\nu}_L=p\widetilde{\nu}+q\widetilde{\nu}^{\ast},
~~~~\widetilde{\nu}_H=p\widetilde{\nu}-q\widetilde{\nu}^{\ast},
\end{eqnarray}
where the parameters $p$ and $q$ are related by
\begin{eqnarray}
\left(\frac{q}{p}\right)^2 =
\frac{\widehat{M}_{12}^{\ast}-\frac{i}{2}\widehat{\Gamma}_{12}^{\ast}}
     {\widehat{M}_{12}-\frac{i}{2}\widehat{\Gamma}_{12}}.
\label{qop}
\end{eqnarray}
It is appropriate for cosmology to consider an initial state at $t=0$ with
equal densities of $\widetilde{\nu}$ and $\widetilde{\nu}^{\ast}$. At
a later time $t$, the state will have evolved into
\begin{eqnarray}
\widetilde{\nu}(t)=g_{+}(t)\widetilde{\nu}(0) +
\frac{q}{p}g_{-}(t)\widetilde{\nu}^{\ast}(0),
~~~~\widetilde{\nu}^{\ast}(t)=\frac{p}{q}g_{-}(t)\widetilde{\nu}(0)
+ g_{+}(t)\widetilde{\nu}^{\ast}(0), \\
g_{+}(t)=e^{-im_{\widetilde{\nu}}t}e^{-\Gamma t/2} \cos(\Delta
m_{\widetilde{\nu}} t/2),~~~~
g_{-}(t)=e^{-im_{\widetilde{\nu}}t}e^{-\Gamma t/2} \sin(\Delta
m_{\widetilde{\nu}} t/2).
\end{eqnarray}
Here $\Delta m_{\widetilde{\nu}} \equiv
m_{\widetilde{\nu_2}}-m_{\widetilde{\nu}_1}$ and we have neglected
$\Delta \Gamma$ with respect to $\Delta m_{\widetilde{\nu}}$. We
can then compute the total integrated lepton asymmetry, defined by
\begin{eqnarray}
\epsilon &=&
\frac{\sum_f\int^{\infty}_{0}dt[\Gamma(\widetilde{\nu}(t)\rightarrow
f)+\Gamma(\widetilde{\nu}^{\ast}(t)\rightarrow
f)-\Gamma(\widetilde{\nu}(t)\rightarrow
\bar{f})-\Gamma(\widetilde{\nu}^{\ast}(t)\rightarrow
\bar{f})]}{\sum_f\int^{\infty}_{0}dt[\Gamma(\widetilde{\nu}(t)\rightarrow
f)+\Gamma(\widetilde{\nu}^{\ast}(t)\rightarrow
f)+\Gamma(\widetilde{\nu}(t)\rightarrow
\bar{f})+\Gamma(\widetilde{\nu}^{\ast}(t)\rightarrow \bar{f})]}\\
&=&\frac{1}{2}\left(\left|\frac{q}{p}\right|^2-\left|\frac{p}{q}\right|^2\right)
\frac{\int^{\infty}_0dt|g_{-}|^2}{\int^{\infty}_0dt(|g_+|^2+|g_-|^2)},
\end{eqnarray}
where $f$ stands for a final state with lepton number equal to one
 and $\bar{f}$ is its conjugate.  Calculating (\ref{qop}) in the limit
$\widehat{\Gamma}_{12}<<\widehat{M}_{12}$, we find
\begin{eqnarray}
\left|\frac{q}{p}\right|^2 &\simeq &
1-\mbox{Im}\left(\frac{\widehat{\Gamma}_{12}}{\widehat{M}_{12}}\right)
\nonumber \\
&\simeq & 1-\mbox{Im} \left(\frac{g^4\tan^4\beta m_l^4
m_{\nu}A_{\nu}I}{512\pi^3M_W^4m^3_{\snu}\Delta m_{\snu}} \right).
\end{eqnarray}
We notice that $A_{\nu}$ is the only complex
parameter in this expression. Performing the time integral, we find
\begin{eqnarray}
\frac{\int^{\infty}_0dt|g_{-}|^2}{\int^{\infty}_0dt(|g_+|^2+|g_-|^2)}
=\frac{(\Delta
m_{\widetilde{\nu}})^2}{2(\Gamma^2+(\Delta
m_{\widetilde{\nu}})^2)}. \label{ofact}
\end{eqnarray}
Thus, we obtain the following final expression for the CP asymmetry:
\begin{eqnarray}
\epsilon &=& -\frac{(\Delta m_{\widetilde{\nu}})^2}{2(\Gamma^2+
(\Delta m_{\widetilde{\nu}})^2)} \frac{g^4\tan^4\beta m_l^4
m_{\nu}\mbox{Im}A_{\nu}I}{512\pi^3M_W^4m^3_{\snu}\Delta m_{\snu}}.
\label{cpasym}
\end{eqnarray}
The baryon asymmetry is then given by \cite{giudice}
\begin{eqnarray}
\frac{n_B}{s}=-\left( \frac{24+4n_H}{66+13n_H} \right)\epsilon
\eta Y^{eq}_{\widetilde{\nu}}.
\end{eqnarray}
The first factor takes into account the reprocessing of the $B-L$
asymmetry by sphaleron transitions, with the number of Higgs
doublets $n_H$ equal to 2, and
$Y^{eq}_{\widetilde{\nu}} = 45\zeta(3)/(\pi^4g_{\ast})$ is the
equilibrium sneutrino density in units of the entropy density for
temperatures much larger than $m_{\widetilde{\nu}}$. Therefore, we
obtain
\begin{eqnarray}
\frac{n_B}{s}=-8.6\times 10^{-4} \epsilon \eta.
\end{eqnarray}
The efficiency factor $\eta$ includes effects caused by
the sneutrino density being smaller than the equilibrium density
and wash-out due to the decays not being completely
out of equilibrium. The value of $\eta$ must be obtained by
integrating the relevant Boltzmann equations, but may plausibly be of
order unity.

For the value of $n_B/s$ given in (1), we see that that our mechanism
should yield $|\epsilon \eta|\simeq 10^{-6}-10^{-7}$. Let us estimate how
we can achieve this desirable amount of $|\epsilon|$. We infer from recent
experimental results that a neutrino mass of order $10^{-2}$ eV is
possible. Then, we see from (5) that $\Delta m_{\widetilde{\nu}}\simeq
10^{-2} \mbox{eV}$ in the case that the parameters $A_N,B_N,\mu$ are of
the same order as $m_{\snu}$. In this case, the size of the quantity in
(\ref{ofact}) is of order one. We can also take the parameter
$\mbox{Im}A_{\nu}$ to be the same order of $m_{\snu}$. Moreover, we
observe that the value of $\Gamma_{12}$ can be enhanced by taking large
$\tan\beta$. Numerically, we find a magnitude of $|\epsilon|$ in
(\ref{cpasym}) of order $10^{-6}-10^{-7}$ when we take $m_{\snu}\simeq 500
\mbox{GeV}, ~m_H\simeq 1 \mbox{TeV}, m_l=m_{\tau}$ and $\tan\beta \simeq
30$, and a larger value of $|\epsilon|$ could be obtained if
needed.

Finally, we note that the large sneutrino-antisneutrino mixing which is
expected because $\Delta m_{\tilde \nu} \gg \Gamma_{\tilde \nu}$ might
lead to an observable like-sign dilepton signal. This could provide an
characteristic collider signature of our scenario. On the other hand, the
amount of CP violation is likely to be too small to be observable in the
near future.

\section{Discussion}

The scenario proposed here needs further investigation.
In particular, we recall that successful leptogenesis requires a
departure from thermal equilibrium for wash-out processes, as well
lepton-number-violating decays. The dominant wash-out processes
are scattering reactions mediated by sneutrinos or charginos and
neutralinos. Since our scenario calls for relatively heavy
charginos and neutralinos, the scattering rates mediated by
charginos and neutralinos are suppressed. Additionally, scattering
reactions mediated by sneutrinos can occur via the vertices
involving charginos or neutralinos. However, the rates of these
scattering processes are suppressed due to the Boltzmann
suppression of the number densities of the charginos and
neutralinos. A more accurate estimation of the lepton asymmetry
that survives these wash-out processes could be obtained by
solving directly the full Boltzmann equations for the system with
different sparticle masses, as has been done in many detailed
works considering high-scale leptogenesis~\cite{giudice}. A
similarly detailed study goes beyond the scope of this paper, and
will be undertaken elsewhere.

In conclusion, an alternative of leptogenesis through decays of a
left-handed sneutrino at a low energy scale has been proposed.
This scenario may be realized in supersymmetric models with non-zero Majorana
masses of neutrino superfield that lead to mixing and mass splitting
between the left-handed sneutrino and the corresponding antisneutrino.
Soft supersymmrty breaking sectors provide new sources of CP violation in
sneutrino-antisneutrino mixing which can generate a lepton asymmetry in
decays of the left-handed sneutrino. We have shown that the three Sakharov
conditions~\cite{sakharov} for the observed baryon asymmetry of the
Universe can be fulfilled in this restrictive framework.

\section{acknowledgement}
SKK is supported in part by BK21 program of the Ministry of
Education in Korea and in part by KOSEF Grant No.
R01-2003-000-10229-0.

\end{document}